\documentclass[aps,pre,twocolumn,floatfix,superscriptaddress,nofootinbib]{revtex4-1}
\usepackage{amsmath, amssymb, graphics, mathrsfs, bm, stackrel,bbold}

\usepackage[usenames,dvipsnames]{xcolor}
\definecolor{Highlight}{rgb}{1,1,0.75}
\newcommand{\authormain}{Srividya Iyer-Biswas}
\newcommand{\titlemain}{Phenomenology of stochastic exponential growth}
\usepackage{graphicx}
\usepackage{epstopdf}
\usepackage[bookmarks={false}, pdfauthor={\authormain}, pdftitle={\titlemain}]{hyperref}
\hypersetup{colorlinks=true, linkcolor=BrickRed, citecolor=Violet, filecolor=OliveGreen, urlcolor=RoyalBlue, filebordercolor={.8 .8 1}, urlbordercolor={.8 .8 0}}
\usepackage[all]{hypcap}

\usepackage{graphicx}   
\usepackage{dcolumn}    
\usepackage{bm}     
\usepackage[caption=false]{subfig}

\usepackage{dsfont}     
\usepackage{color}		

\newcommand\der[2]{\frac{\text{d}{#1}}{\text{d}{#2}}}

\newcommand\inv[1]{\frac{1}{#1}}

\newcommand\e{\text{e}}
\newcommand\ie{\textit{i.e.}~}

\newcommand\eq[1]{Eq.~(\ref{eq:#1})}
\newcommand\Eq[1]{Equation~(\ref{eq:#1})}
\newcommand\fig[1]{Fig.~\ref{fig:#1}}
\newcommand\Fig[1]{Figure~\ref{fig:#1}}

\newcommand\sect[1]{Section~\ref{sec:#1}}
\newcommand\app[1]{Appendix~\ref{app:#1}}
\newcommand\rf[1]{Ref.~\cite{#1}}

\newcommand{\mean}[1]{\left\langle#1\right\rangle}
\newcommand{\var}[1]{\text{var}\left[#1\right]}

\newcommand{\red}[1]{\textcolor{red}{#1}}

\usepackage[normalem]{ulem}

\newcommand{\fj}[1]{{\color{blue} #1}}

\newcommand\ba{\begin{array}}
\newcommand\ea{\end{array}}
\newcommand\nn{\nonumber}

\newcommand\ri{\right}
\renewcommand\le{\left}

\newcommand{\feyn}[1]{#1\kern-0.45em/}

\newcommand{\tto}{\rightarrow}

\newcommand\g{\gamma}

\renewcommand\k{\kappa}

\newcommand\m{\mu}

\begin{document}
\title{Phenomenology of stochastic exponential growth}
\author{Dan Pirjol}
\affiliation{National Institute of Physics and Nuclear Engineering, Bucharest, Romania}
\author{Farshid Jafarpour}
\affiliation{Department of Physics and Astronomy, Purdue University, West Lafayette, IN 47907}
\author{Srividya Iyer-Biswas}
\email{iyerbiswas@purdue.edu}
\affiliation{Department of Physics and Astronomy, Purdue University, West Lafayette, IN 47907}

\begin{abstract}
Stochastic exponential growth is observed in a variety of contexts, 
including molecular autocatalysis, nuclear fission, population growth, 
inflation of the universe, viral social media posts, and financial markets. 
Yet literature on modeling the phenomenology of these stochastic dynamics 
has predominantly focused on one model, Geometric Brownian Motion (GBM), 
which can be described as the solution of a Langevin equation with linear drift and linear multiplicative noise. Using recent experimental results on stochastic exponential growth of 
individual bacterial cell sizes, we motivate the need for a more general 
class of phenomenological models of stochastic exponential growth, which are 
consistent with the observation that the mean-rescaled distributions are 
approximately stationary at long times. We show that this behavior is not 
consistent with GBM, instead it is consistent with power law multiplicative 
noise with positive fractional powers. Therefore, we consider this general 
class of phenomenological models for stochastic exponential growth, provide 
analytical solutions, and identify the important dimensionless combination 
of model parameters which determines the shape of the mean-rescaled 
distribution. We also provide a prescription for robustly inferring model 
parameters from experimentally observed stochastic growth trajectories.
\end{abstract}

\pacs{\red{???,???,\dots}}
\maketitle

\section{Introduction}

Stochastic exponential growth occurs when a quantity of interest multiplies 
geometrically, with a random multiplier and/or with stochastic waiting times between growth spurts. 
Such processes have been observed in various contexts, across a range of 
length and timescales. Familiar examples include microbial population 
growth~\cite{ginzburg1982quasiextinction, keiding1975extinction,LewCohen, 2016-iyerbiswas-arxiv}, 
nuclear fission~\cite{EverettUlam1948,BirkhoffVarga1958}, 
copy number dynamics of key 
biochemicals~\cite{iyer2014universality, tsuru2009noisy, friedman2006linking, cai2006stochastic},  
increase of individual cell sizes during interdivision 
periods~\cite{iyer2014scaling, iyer2014universality, pfeuty2007minimal, 2016-iyerbiswas-fpt}, tissue growth in 
developmental and cancer biology~\cite{eden1961two, furusawa2000complex, kohandel2007dynamics, halpin1995kinetic}, 
spreading of epidemics~\cite{girvan2002simple},  growth of internet and 
propagation of viral social media posts~\cite{huberman1999internet, kumar2000stochastic}, 
financial markets~\cite{samuelson1965,black1973pricing}, stochastic divergence of neighboring phase 
space trajectories~\cite{stoddard1973numerical}, and  cosmological expansion of the 
universe~\cite{diamandis1999ongraceful}.

Despite the variety of contexts in which the phenomenon occurs, literature on modeling 
stochastic exponential growth has predominantly focused on one model, Geometric Brownian Motion (GBM). Such processes appear naturally in the context of
proportionate growth processes; for a general introduction to these processes and their 
applications see \cite{Mitzenmacher2004}. A stochastic process, $x(t)$, is said to undergo GBM if its time evolution obeys a Langevin equation with linear drift and linear multiplicative noise~\cite{VanKampen}:

\begin{equation}
	\der{x}{t} = \kappa\, x + \sqrt{D} \, x\, \xi(t),
\end{equation}
in which $\xi(t)$ is Gaussian white noise, traditionally interpreted in the 
It\^o sense~\cite{VanKampen}, with 
\begin{equation}
	\mean{\xi(t)} = 0, \text{\;and\;} \mean{\xi(t)\xi(t')} = \delta(t-t').
\end{equation}
It is well-known that this process yields a log-normal distribution for $x$, and the ratio of its 
standard deviation to mean diverges with time~{\cite{VanKampen}}.

Recent studies have revealed a scaling law governing fluctuations during stochastic exponential growth of individual microbial cell sizes in between divisions: the mean-rescaled distribution is stationary and has finite variance at long times~\cite{iyer2014universality,iyer2014scaling}. This observation is inconsistent with the behavior predicted from GBM, in which the coefficient of variation (ratio of standard deviation to mean) diverges with time, and so the mean-rescaled distribution is not stationary~{\cite{VanKampen}}.

Motivated by these observations, we seek a general phenomenological description 
of stochastic exponential growth, consistent with the time independence of the 
mean-rescaled distribution at long times. To this end we consider a 
{stochastic Markovian growth process which is a}
generalization of GBM, 
in which the multiplicative noise term is given by a power-law with exponent 
$\g$:
\begin{equation} \label{eq:gamma}
	\der{x}{t} = \kappa\, x + \sqrt{D}\, x^\gamma \xi(t).
\end{equation}
{This model is well-studied in finance literature in the context of applications in option pricing~\cite{cox1996constant, lo2000constant}. }We find  constraints on $\gamma$ for which these dynamics are consistent with stationarity 
of the asymptotic mean-rescaled distribution. We show that the probability density of 
the mean-rescaled variable, $y(t) \equiv x(t)/\mean{x(t)}$, asymptotically approaches a 
time-independent distribution with finite variance provided $0\leq\gamma<1$. 
Thus we focus on the phenomenology of stochastic exponential growth for the class of models 
given by \eq{gamma}, with  $0\leq\gamma<1$. For this class of models we provide analytical 
solutions for the time dependent dynamics, and use them to motivate a robust prescription 
for inferring model parameters  ($\gamma$, $\kappa$\fj{,} and $D$) from observed stochastic 
growth trajectories.

In \sect{fvmrd} we comment on generic properties of a more general class of models than
\eq{gamma}, namely, ones with a general form of the multiplicative noise term. Remarkably, 
a stable distribution for the mean-rescaled variable, $y(t)$, is found to exist in the long 
time limit for this general class of models. We derive a sufficient condition for finiteness 
of the variance of this asymptotic distribution, and use this constraint to motivate why we 
focus on the phenomenology of the class of models in \eq{gamma}. In Section III, we discuss 
analytical solutions and general properties of \eq{gamma}. We also examine in detail the 
limiting cases corresponding to $\g = 0$ and $\g = 1$, and a special case, $\g = 1/2$, which 
arises naturally from a microscopic model for stochastic exponential growth~\cite{iyer2014universality}. 
Using the analytical solutions for \eq{gamma}, we find that a specific combination of the second 
and third central moments of the distributions is particularly sensitive to the exponent $\gamma$, 
and not to other model parameters. In Section IV we utilize this idea to provide a robust 
prescription for inferring the exponent $\gamma$ from   time series data, and test the prescription 
using synthetic data from numerical simulations of \eq{gamma}. Once $\g$ has been determined, 
estimation of the parameters $\kappa$ and $D$ is a straightforward statistical inference problem. 
Here we use the Maximization of Likelihood Estimation (MLE) approach of parametric statistical 
regression, and present its application to the models considered.


\section{Condition for Finite Variance of the Mean-Rescaled Distribution} \label{sec:fvmrd}
To begin with, we consider properties of the mean-rescaled distribution for a general class of 
stochastic exponential growth models with multiplicative noise:
\begin{eqnarray}\label{eq:sde}
	\der{x}{t} = \kappa\, x + f(x) \xi(t).
\end{eqnarray}
We will assume the initial condition, $x(t=0) = x_0$, and an absorbing boundary at the origin, $x=0$, 
to ensure that $x(t)$ remains non-negative.  The phenomenological model introduced in \eq{gamma} 
corresponds to the subset of models with multiplicative noise, $f(x)$, of the form $\sqrt{D} x^{\g}$.

In the absence of noise\fj{,} $x$ grows exponentially with time: $x(t) = x_0\, e^{\kappa t}$. 
When fluctuations are present, the ensemble averaged value undergoes exponential growth with growth rate $\k$:
\begin{eqnarray}\label{eq:mean}
	\mean{x(t)} = x_0\, e^{\kappa t}.
\end{eqnarray}
This follows directly from \eq{sde} upon taking the expectation value and 
integrating over time. We remind the reader that \eq{sde} is to be interpreted 
in the It\^o sense.

What can be said about the spread of stochastic trajectories, $x(t)$, around 
the expectation value given in \eq{mean}, in the presence of the noise term 
in \eq{sde}? We are especially interested in the 
mean-rescaled variable $y(t) \equiv x(t)/\mean{x(t)}$, and its behavior in 
the long time limit, $t\to \infty$. Under general conditions\footnote{Assuming
that the noise term $f(x)$ satisfies appropriate regularity conditions which
are required for the existence and uniqueness of the solution of the stochastic 
differential equation \eq{sde},
it can be shown that $y(t)$ is a continuous time, non-negative martingale.
It is known that any non-negative supermartingale (martingale)
is $L^1$ integrable. By the
Doob's martingale convergence theorem [37], it follows that the
limit $\lim_{t\to \infty} y(t) = y_\infty$ exists, in an almost sure sense.
Furthermore, $y_\infty$ is also $L^1$ integrable. 
This implies that $y(t)$ converges to a well-defined random variable in
long time limit.},
$y(t)$ approaches a limiting distribution in the long time limit. However, 
this result does not place constraints on the {\em finiteness} of the variance 
of the asymptotic distribution of $y(t)$. Motivated by experimental 
observations in  \cite{iyer2014scaling,iyer2014universality}, 
we are interested in possible functional forms of the noise 
term $f(x)$, for which the distribution of $y(t)$ has a finite variance as 
$t\to \infty$. 

To this end, we first find an expression for the variance of $y(t)$. Changing 
variables in \eq{sde},
\begin{equation}
\der{y}{t} =x_0^{-1} e^{-\kappa t} f\left(y\, x_0 e^{\kappa t}\right) \xi(t).
\end{equation}
Using It\^o\rq{}s lemma~{\cite{McKean}},
\begin{equation} \label{eq:dy^2}
\begin{split}
	\der{y^2}{t} =\, &x_0^{-2} e^{-2\kappa t} f^2\left(y\,x_0 e^{\kappa t}\right)\\
		&+2\,y\,x_0^{-1}\,e^{-\kappa t} f\left(y\, x_0 e^{\kappa t}\right) \xi(t).
\end{split}
\end{equation}
The variance, $\var{y(t)}$, becomes equal to $\mean{y(t)^2} - 1$, 
since $y(t)$ is a mean-rescaled variable {and its average is 
$\mean{y(t)}=1$}. Thus, using \eq{dy^2}, we have
\begin{equation}
	\der{}{t} \var{y(t)} =x_0^{-2} e^{-2\kappa t} {\mean{f^2\left(y\,x_0 e^{\kappa t}\right)}},
\end{equation}
which implies
\begin{equation}
	\var{y(t)} =  x_0^{-2} \int_0^t dt\rq{}\, e^{-2\kappa t\rq{}}\mean{f^2(x(t\rq{}))}.
\end{equation}
Therefore, the mean rescaled distribution has a finite variance only if
\begin{equation}\label{eq:cond}
	\int_0^\infty dt\, e^{-2\kappa t}\mean{f^2(x(t))} < \infty.
\end{equation}

What are the possible functional forms of $f(x)$ which satisfy the 
condition \eq{cond}? Evidently, any bounded function $f(x) < M$ will 
satisfy this condition. Thus, any model of multiplicative noise for which the amplitude of the noise saturates for large $x$ will produce a finite variance distribution for $y$ in the large time limit. 

{\em The small-noise limit.}
In this limit, $f(x) \ll \sqrt{\kappa} \, x$, we can approximate the expectation with
the deterministic value,
\begin{eqnarray}\label{c2}
	\mean{f^2(x)} \simeq f^2(x_0 e^{{\kappa}t}).
\end{eqnarray}
Therefore the condition on $f(x)$, for finiteness of variance of the asymptotic mean-rescaled distribution, can be simply stated  as $f^2(x)/x^2$ must 
fall off faster than $1/\log x$ in the limit $x\to \infty$, such that the integral in 
\eq{cond} converges as $t \to \infty$. This analysis suggests a more explicit
condition on $f(x)$:
\begin{eqnarray}\label{eq:smallnoise}
	\lim_{x\to \infty} \log x \frac{f^2(x)}{x^2} = 0.
\end{eqnarray}
Thus if the leading behavior of $f(x)$ in the large $x$ limit is a power law, i.e., $f(x \tto \infty) \sim x^{\g}$, then the mean-rescaled distribution can become stationary only for $0 \leq \g < 1$. Hence our focus on the phenomenological models of stochastic exponential growth given by \eq{gamma}.

In the following Section, before proceeding to give a general solution for 
models of the form in \eq{gamma}, we will contextualize the results by 
carefully considering the limiting cases, $\g = 0$ (constant additive noise) 
and $\g=1$ (linear multiplicative noise, corresponding to GBM). 
In addition, we will also consider the $\g = 1/2$ (square-root multiplicative 
noise) case in detail, {since it arises naturally in discrete stochastic models (for e.g., the stochastic Hinshelwood cycle~\cite{iyer2014universality}, birth-death and Galton-Watson branching processes \cite{EthierKurtz}).}

In relation to \eq{smallnoise}, the expected behavior of the asymptotic 
mean-rescaled distributions for these cases are summarized below.

\begin{itemize}
	\item Constant additive noise, $f(x) = \sqrt{D}$:  since the noise
has bounded intensity for this case, the asymptotic distribution of $y(t)$ has finite variance. This distribution can be found exactly and is Gaussian (see \sect{power_law}).
	\item Linear multiplicative noise, $f(x) = \sqrt{D} x$:
this corresponds to geometric Brownian motion (GBM). Here the condition
\eq{smallnoise} is not satisfied, and thus, as explicitly derived in the following Section, the variance of $y(t)$ increases without limit.
	\item Power-law multiplicative noise, $f(x) = \sqrt{D} x^\gamma$, with 
$0 < \gamma < 1$: since the condition in \eq{smallnoise} is satisfied for this 
choice of $f(x)$, the variance of the asymptotic mean-rescaled distribution is 
finite. We will show this explicitly in \sect{power_law}.
\end{itemize}

\section{Power-law multiplicative noise: solutions and properties} \label{sec:power_law}

As motivated in the preceding section, we now focus on the phenomenology and analytical solutions of stochastic exponential growth process with power-law multiplicative noise:
\begin{equation}
	\der{x}{t} = \kappa\, x + \sqrt{D}\, x^\gamma \xi(t).
	\tag{\ref{eq:gamma} revisited}
\end{equation}
To contextualize the general results, we first review the properties of the 
familiar special cases of this equation, i.e., $\gamma \in \{0, 1, 1/2\}$ in 
\sect{power_law}A --- C, before proceeding to the general solution in 
\sect{power_law}D. The constant additive noise case, $\gamma = 0$, corresponds 
to a simple stochastic differential equation which is known to have a 
Gaussian solution~\cite{OUprocess,VanKampen}. 
Linear multiplicative noise, $\gamma = 1$, 
corresponds to geometric Brownian motion (GBM). This process was proposed 
as a model for risky asset prices by Samuelson \cite{samuelson1965}, and was 
used by Black and Scholes in their pricing theory for equity derivatives 
~{\cite{black1973pricing}}. {Square-root multiplicative noise, $\gamma = 1/2$, corresponds to the 
square-root process~\cite{feller1951two}; this is obtained as the continuous limit of discrete
stochastic models such as the the stochastic Hinshelwood cycle~\cite{iyer2014universality}. More generally, square-root processes arise in the context of birth-death and Galton-Watson branching processes~{\cite{EthierKurtz, 2009-iyerbiswas, 2014-iyerbiswas-pre, 2009-iyerbiswas-dissertation}}.}

\subsection{Constant additive noise, $\gamma = 0$}

The simplest model of stochastic exponential growth has constant amplitude
noise, $f(x)=\sqrt{D}$, and the corresponding stochastic differential equation has a simple Gaussian noise term:
\begin{eqnarray}\label{eq:constant}
	\der{x}{t} = \kappa\, x + \sqrt{D} \, \xi(t).
\end{eqnarray}
This immediately suggests that the dimensionless ratio,  
\begin{equation}\label{eq:sigma_0}
	\Sigma_{0}^2 \equiv \frac{D}{2 \kappa\, x_0^2},
\end{equation} 
which can be identified as the variance of the mean-rescaled distribution, should be a key 
determinant of its shape (the subscript $0$ for $\Sigma^{2}$ indicates that this corresponds 
to $\gamma = 0$). {This dimensionless variable is the inverse P\'eclet number 
of the process, 
and as we will see, we will need to generalize the P\'eclet number (or its 
inverse) for 
different values of $\gamma$.}

In the absence of a boundary condition at the origin, 
this process {is a mean-repelling Ornstein-Uhlenbeck process~\cite{OUprocess}.
We remind the reader that the Ornstein-Uhlenbeck process has the form
${dx}/{dt}=\kappa (\m - x) + \sqrt{D} \xi(t)$, where $\kappa >0$ 
corresponds to the mean-reverting case, and $\kappa<0$ is the mean-repelling case.
For both cases this is known to yield a Gaussian distribution.}
Therefore, the mean-rescaled distribution, $q(y, t)$, is also Gaussian, with 
variance given by
\begin{equation}
	\var{y(t)} = \Sigma_{0}^2 \,\tau(t),
\end{equation}
with 
\begin{equation}
	\tau(t) \equiv 1 - e^{-2\kappa t};
\end{equation} 
$\tau(t)$ identifies the characteristic timescale of approaching  stationary-state. Thus, the mean-rescaled distribution, $q(y, t)$,  asymptotically approaches a time-independent Gaussian distribution with variance $\Sigma_{0}^2$, for $t \gg 1/(2\k)$. By definition, the mean of the distribution is unity at all times.
 
In proximity of the origin this stochastic {process} can become negative. 
However, in real-world applications of the stochastic exponential growth 
process (e.g., the growth dynamics of bacterial cell sizes), the dynamical 
variable (e.g., the cell's size) may be constrained to always remain positive. 
This can be addressed in the model by imposing an absorbing boundary condition 
at the origin, $x = 0$, to avoid negative values for the dynamical variable. 
We can compute the solution for $p(x, t|x_0, 0)$, the probability density of 
$x$ at time $t$ given the initial condition $x_0$ at time $t = 0$, by using 
the method of images (see \app{constant_noise} for details of derivation).
We obtain
\begin{equation}
	p(x, t|x_0, 0) = \frac{1}{x_0} e^{-\kappa t} q_{\rm abs}
		\left( \frac{1}{x_0} e^{-\kappa t} x, \tau(t)\right),
\end{equation}
where 
\begin{equation}\label{eq:qGauss}
\begin{split}
	q_{\rm abs}(y,\tau) \equiv& \, \frac{2}{\sqrt{2\pi\tau(t)} \, \Sigma_{0}} \exp\left[ - \frac{y^2+1}{2\,\Sigma_{0}^2 \, \tau(t)} \right]\\
		 &\times \sinh\left[ \frac{y}{\Sigma_{0}^2 \,\tau(t)}  \right].
\end{split}
\end{equation}

Since the modified time, $\tau(t)$, approaches a finite limit ($= 1$) in 
the long time limit, it follows
that $y(t)$ approaches a stationary distribution asymptotically; this distribution is  
given by using $\tau(t) \to 1$ in \eq{qGauss}.

The probability of absorption at origin is 
\begin{eqnarray}
	p_{\rm abs}(t) &=& 1 - \int_0^\infty dy \, q_{\rm abs}(y,t) \\
&=& 1 - \mbox{Erf}\left[ \frac{1}{\sqrt{2 \, \Sigma_0^2 \,\tau(t)}} \right].\nonumber
\end{eqnarray}

The mean of $y(t)$ is equal to $1$, as expected, and the second moment is 
given by
\begin{eqnarray}
\begin{split}
	\mean{y^2(t)} =& (1+v(t))\,\mbox{Erf}\left[\frac{1}{\sqrt{2v(t)}}\right]\\
		&+ \sqrt{\frac{2v(t)}{\pi}}\exp \le[-\frac{1}{2v(t)}\ri],
\end{split}
\end{eqnarray}
where $v(t) \equiv \Sigma_0^2\,\tau(t)$. 
For $v(t) \ll 1$, one has $\var{y(t)} \simeq v(t)$. This variance approaches a limiting value in the infinite time limit:  $\var{y(t)} \simeq {D}/({2\kappa \,x_0^2}) \equiv \Sigma_0^2$.


\subsection{Linear multiplicative noise (geometric Brownian Motion), $\gamma = 1$}

The case of $\g = 1$ corresponds to another simple model. Here the noise amplitude is proportional to the magnitude of the process, $x(t)$ and so
\begin{eqnarray}
	\der{x}{t} = \kappa\, x + \sqrt{D}\, x\, \xi(t).
\end{eqnarray}
This process has been well studied and is known as geometric Brownian motion (GBM) 
{\cite{VanKampen}}.
The solution for $x(t)$ is
\begin{eqnarray}
	x(t) = x_0 e^{\sqrt{D} \, W(t) + (\kappa - D/2) t},
\end{eqnarray}
where $W(t)$ denotes standard Brownian motion.

Using It\^o\rq{}s lemma, we find that the mean-rescaled process also follows geometric Brownian motion:
\begin{eqnarray}
	y(t) = e^{\sqrt{D} W(t)  - \frac12 D t} \,.
\end{eqnarray}
The probability density of the mean-rescaled variable, $q(y, t)$, is given by 
\begin{equation}
	q(y, t) = \inv{y\sqrt{2\pi D t}} \exp\left[{-}\frac{\left(\ln(y)+\inv 2 D \,t\right)^2}{2D \,t}\right].
\end{equation}

The variance of $y(t)$ is given by 
\begin{eqnarray}
&& \var{y(t)} = e^{D \,  t}-1,
\end{eqnarray}
while the mean is always unity.

In the small time limit, $t \ll 1/D$,  the variance increases linearly as $\var{y(t)}
\sim t$, while in long time limit, $t \gg 1/D$, it increases  
exponentially as $\var{y(t)} \sim e^{D t}$. Thus, for $\g = 1$, the variance
of $y(t)$ does not approach a finite asymptotic value in the infinite time
limit. The distribution of $y(t)$ becomes 
increasingly concentrated near zero, as can be seen by computing the probability
that $y(t)$ is larger than an arbitrary value, $\epsilon$.
\begin{eqnarray}
&& \mathbb{P}(y(t)>\epsilon) = \int_\epsilon^\infty q(y,t)dy \nn \\
&& = \int_{\log\epsilon+\frac12 Dt}^\infty \frac{dz}{\sqrt{2\pi Dt}} e^{-\frac{1}{2Dt}z^2}\nn\\
&& = \Phi\left( - \frac{1}{\sqrt{Dt}} (\log\epsilon+\frac12 Dt) \right) \,.
\end{eqnarray}
Here $\Phi(z)=\int_z^\infty \frac{dt}{\sqrt{2\pi}} e^{-\frac12 t^2}$
is the normal cumulative distribution function. This probability goes to zero as $t\to \infty$, for any $\epsilon>0$. At the same time, the variance of $y(t)$ increases without limit. The seemingly paradoxical conclusion that the distribution of $y(t)$ becomes more and more concentrated near the origin while its variance grows without limit can be understood using the argument provided by Lewontin and Cohen \cite{LewCohen},  who formulated it in the general context of the proportionate growth models~\cite{Mitzenmacher2004}.


\subsection{Square-root multiplicative noise, $\gamma = \frac12$}

Next we consider the special case, $\gamma=\frac12$, which arises naturally in the context of phenomenological models corresponding to discrete  stochastic exponential growth process~\cite{feller2008introduction}. Rewriting \eq{gamma} with a square-root multiplicative noise term, one has 
\begin{eqnarray}\label{eq:sqrt}
	\der{x}{t} = \kappa\, x + \sqrt{D \,x}\, \xi(t).
\end{eqnarray}
Processes with multiplicative square-root noise often appear in models of population genetics and in the Langevin
description of birth-death processes~\cite{feller2008introduction}. The properties of 
the stochastic differential equations with multiplicative square-root noise have been studied in detail by 
Feller in \rf{feller1951two}.

In the absence of the noise term this process describes exponential growth
with a constant growth rate, $\kappa>0$. The noise term introduces fluctuations which
have variance proportional to $x$. The transition density
$p(x,t|x_0,0)$ can be read off from \rf{feller1951two} (also see Equation (1.6) in
\rf{mazzon2012processo}), and is given by
\begin{equation}\label{eq:fsqrt}
\begin{split}
	p(x,t|x_0;0) =& \frac{\kappa}{D \sinh(\kappa \,t/2)}
\exp\left[ - \frac{2\kappa (x+x_0 e^{\kappa \, t})}{D (e^{\kappa\, t}-1)}\right] \\
		&\times \sqrt{\frac{x_0}{x}} I_1 \left( \frac{2\kappa}{D \sinh(\kappa \, t/2)} 
\sqrt{x_0 x} \right)\,,
\end{split}
\end{equation}
where $I_1(x)$ is the modified Bessel function of the first kind.

One can check by direct calculation that the transition density 
satisfies the following relations.
\begin{eqnarray}\label{eq:norm}
& & \int_0^\infty dx\, p(x,t|x_0,0) = 1 - 
    \exp\left[ - \frac{2\kappa x_0}{D(1 - e^{-\kappa t})} \right].\\
\label{eq:1stmom}
& & \int_0^\infty dx\, x\, p(x,t|x_0,0) = \mean{x(t)} = x_0 e^{\kappa t}.
\end{eqnarray}
The relation \eq{norm} encodes normalization of the probability while accounting for the non-zero probability of absorption at the boundary. As expected, in \eq{1stmom} we find that the mean grows exponentially with time, with growth rate, $\k$.

Using It\^o\rq{}s lemma, we find that the second moment, $\mean{x^2(t)}$, satisfies the following time evolution equation:
\begin{eqnarray}
	\der{}{t} \mean{x^2(t)} = 2\kappa\, \mean{x^2(t)} + D x_0\, e^{\kappa t}.
\end{eqnarray}
Solving it with initial condition, $\mean{x^2(0)} \equiv x_0^2$, one has
\begin{eqnarray}\label{eq:2ndmom}
	\mean{x^2(t)} = x_0^2 e^{2\kappa t} {\left[1+ \frac{D}{\kappa x_0}
\left( 1 - e^{-\kappa t} \right)\right].}
\end{eqnarray}
Thus the variance of $x(t)$ is
\begin{eqnarray}\label{xtvar}
	\var{x(t)} = {x_0^2 e^{2\kappa t}\left[\frac{D}{\kappa x_0}
\left( 1 - e^{-\kappa t} \right)\right].}
\end{eqnarray}
Once again, we can identify a dimensionless variance, a combination of parameters which determines the shape of the mean rescaled distribution, from the preceding equation:
\begin{equation}
	\Sigma_{\frac{1}{2}}^2 \equiv \frac{D}{\kappa {x_0}}.
\end{equation}

We can find the density of the mean rescaled variable $y(t)$ directly from the distribution of $x(t)$ given in \eq{fsqrt}. We present next an alternative approach, based on the method
of the time change. This has the advantage that it is applicable also to the 
case of general exponent $\gamma$, which will be discussed next.

An application of It\^o's lemma yields the following stochastic
differential equation for $y(t)$:
\begin{eqnarray}\label{eq:ysde}
	\der{y}{t} = \sqrt{\frac{D\, y}{x_0} } e^{-\frac12 \kappa t} \,\xi(t), 
	\end{eqnarray}
with the initial condition $y_0=1$. 
The time-dependent factor can be absorbed into a deterministic rescaling of time. To this end, we define the modified dimensionless time, $\tau(t)$, by
\begin{eqnarray}
\tau(t) \equiv 1-e^{-\kappa t},
\end{eqnarray}
which implies $d\tau = \e^{-\kappa t}\kappa\, dt$. In terms of $\tau$, \eq{ysde} can be written as
\begin{eqnarray}
	\der{y}{\tau} = \Sigma_{\frac{1}{2}}\sqrt{y}\, \xi(\tau).
\end{eqnarray}
The probability density of $y(t)$ can be read off from \eq{fsqrt}:
\begin{eqnarray}\label{eq:pasympt}
	q(y,t) = 
		\frac{2}{v(t)\sqrt{y}} \exp \le[{-\frac{2 (1+y)}{v(t)}}\ri]
	I_1\left(\frac{4\sqrt{y}}{v(t)}\right)\,,
\end{eqnarray}
where
\begin{equation}\label{eq:v(t)}
	v(t) = \Sigma_{\frac{1}{2}}^{2} \tau(t).
\end{equation}
The variance of $y(t)$ is given by $\var{y(t)} = v(t)$. As $t \to \infty$ 
limit, $\tau$ approaches $1$, thus the mean-rescaled distribution tends to a 
stationary distribution with variance equal to $\Sigma_{\frac{1}{2}}^{2} $.

In the small noise limit, i.e., $D \ll \kappa\,x_0$, the asymptotic probability density, $q^{*}(y)$, is sharply peaked around $1$, and  is well approximated as
\begin{eqnarray}
	q^{*}(y) \simeq \frac{1}{\sqrt{2\pi} \,\Sigma_{\frac{1}{2}}\, y^{3/4} }
\exp\left[-\frac{2(\sqrt{y}-1)^2}{\Sigma_{\frac{1}{2}}^2} \right].
\end{eqnarray}
This relation follows from the general result, \eq{pasympt}, upon using the 
asymptotic expansion of the 
Bessel function, $I_\alpha(x)$: in the limit $x\to \infty$, it approaches 
$\frac{e^x}{\sqrt{2\pi x}} (1 + O(x^{-1}))$.

\subsection{General exponent $\gamma$}
\begin{figure}[t]
\begin{center}
\includegraphics[width=0.48\textwidth]{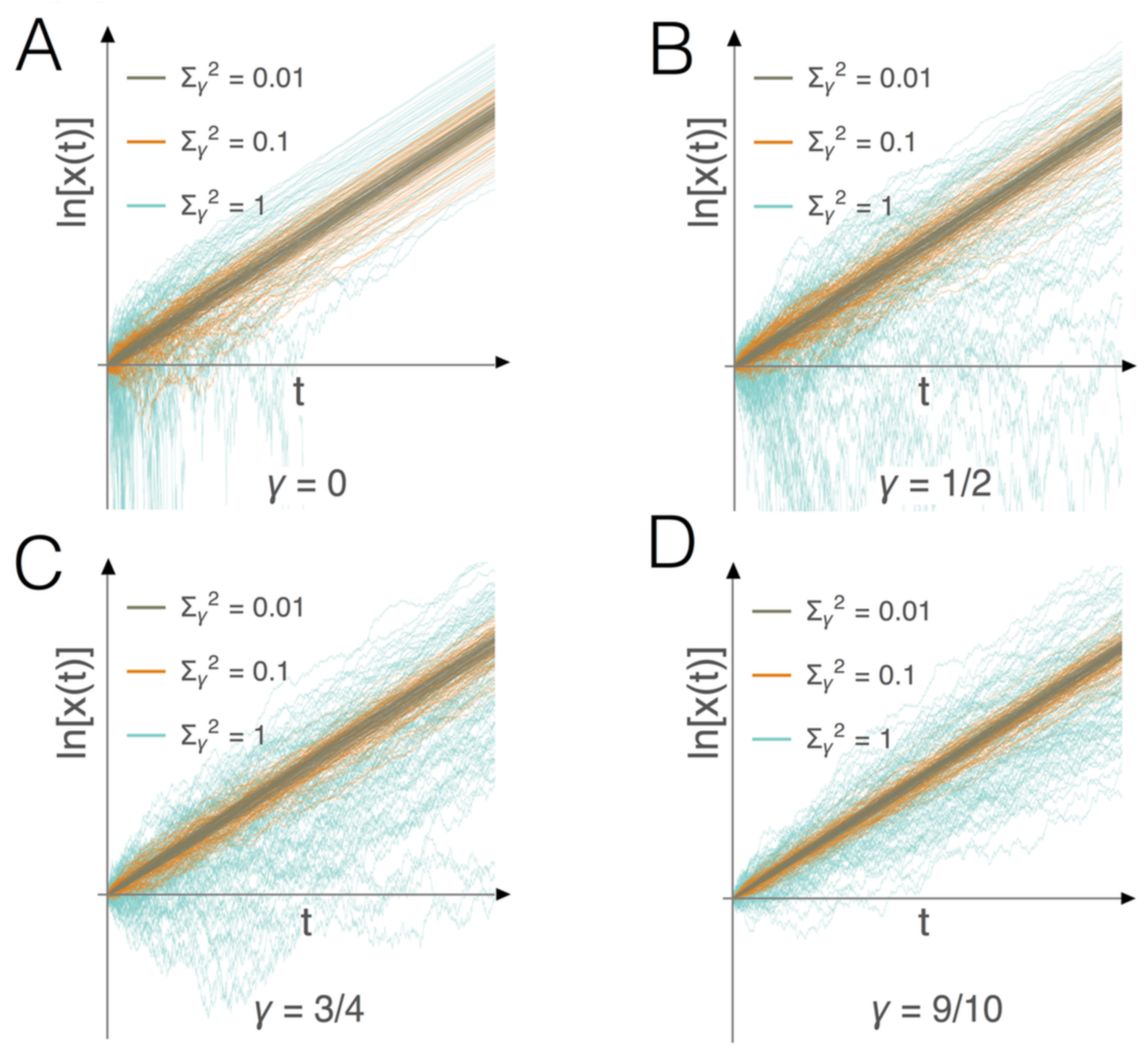}
\caption{Log-linear plot of ensemble trajectories of stochastic exponential growth for different values of the two dimensionless parameters that affect the distributions, namely $\gamma$ and $\Sigma_\gamma^2$. The values are chosen from $\gamma \in \{0,\, 1/2,\, 3/4,\, 0.9\}$ and $\Sigma_\gamma^2 \in \{0.01,\, 0.1,\, 1\}$. The trajectories corresponding to $\Sigma_\gamma^2 =1$ are shown in cyan (light gray), $\Sigma_\gamma^2= 0.1$ in orange (medium gray) and $\Sigma_\gamma^2 = 0.01$ in gray (dark gray). For small values of $\gamma$, the trajectories are more stochastic at the beginning and more likely to be absorbed by the boundary, but become more or less deterministic when they grow large. In contrast, as $\gamma$ approaches $1$, the trajectories stay stochastic for longer time, as the mean rescale distribution takes longer and longer to settle to a stationary distribution. For a fixed $\gamma$, the over all variance grows with $\Sigma_\gamma^2$. For a comparison of the effect of $\gamma$ on the distribution, when $\Sigma_\gamma^2$ kept constant, see \fig{qgamma}. Simulation variables: $x_0 = 1$ and $\kappa = 1$.}
\label{fig:trajectories}
\end{center}
\end{figure}

We have seen that for $\gamma = 0$ and $1/2$, there are characteristic 
timescales equal to ${1/(2\kappa)}$ and ${1}/{\kappa}$, respectively, after which the 
mean-rescaled distributions become stationary. After this time, the shape 
of the distributions only depends on the dimensionless variables $\Sigma_0$ 
and $\Sigma_{1/2}$. In this section, we first find the corresponding 
timescale for a general $\gamma < 1$. Then, we find a dimensionless variable, 
$\Sigma_\gamma$, that determines the shape of the mean rescaled distribution 
at steady state. This dimensionless variable is a generalization of the 
inverse P\'eclet number, which is often used in 
transport literature~{\cite{PecletBook}}.

Let us start with the case of a power-law multiplicative noise  of the form $f(x) = \sqrt{D} x^\gamma$,
with a general exponent, $\gamma > 0$.
\begin{equation}
	\der{x}{t} = \kappa\, x + \sqrt{D} \, x^\gamma\, \xi(t).
	\tag{\ref{eq:gamma} revisited}
\end{equation}
This process appears in financial literature in the context of models 
known as Constant Elasticity of Variance (CEV)~\cite{cox1996constant}. 
We will show that its properties can be understood by relating it to a 
generalization of the square-root process ($\gamma = 1/2$) that we considered 
in the preceding section.
The small-noise approximation result, \eq{smallnoise}, suggests that $\gamma$ 
must be restricted to the range $\gamma < 1$, in order to have a finite 
variance for
$y(t)$ in the large time limit. We will prove this constraint rigorously.

We would like to derive the probability density, $p(x,t|x_0,0)$, for the process defined by \eq{gamma}. It will prove more convenient to first derive the probability density
of the mean-rescaled variable, $q(y,t)$, and subsequently obtain $p(x,t|x_0,0)$ using
\begin{eqnarray}\label{eq:pgamma}
	p(x, t| x_0, 0) = \frac{1}{x_0\,e^{\kappa t}}\, 
	q\left(\frac{x}{x_0\,e^{\kappa t}}, t\right).
\end{eqnarray}
Note that the properties of the mean rescaled distribution should depend on the two dimensionless parameters $\gamma$ and $D/(\kappa \, x_0^{2(1-\gamma)})$. It would however simplify the algebra if we define the dimensionless variance as the following combination of these parameters:
\begin{equation}
	\Sigma_\gamma^2 = \frac{D}{2 (1-\gamma) \kappa\, x_0^{2(1-\gamma)}}.
\end{equation}

\begin{figure*}[t!]
\begin{center}
\resizebox{
0.99\textwidth}{!}{\includegraphics{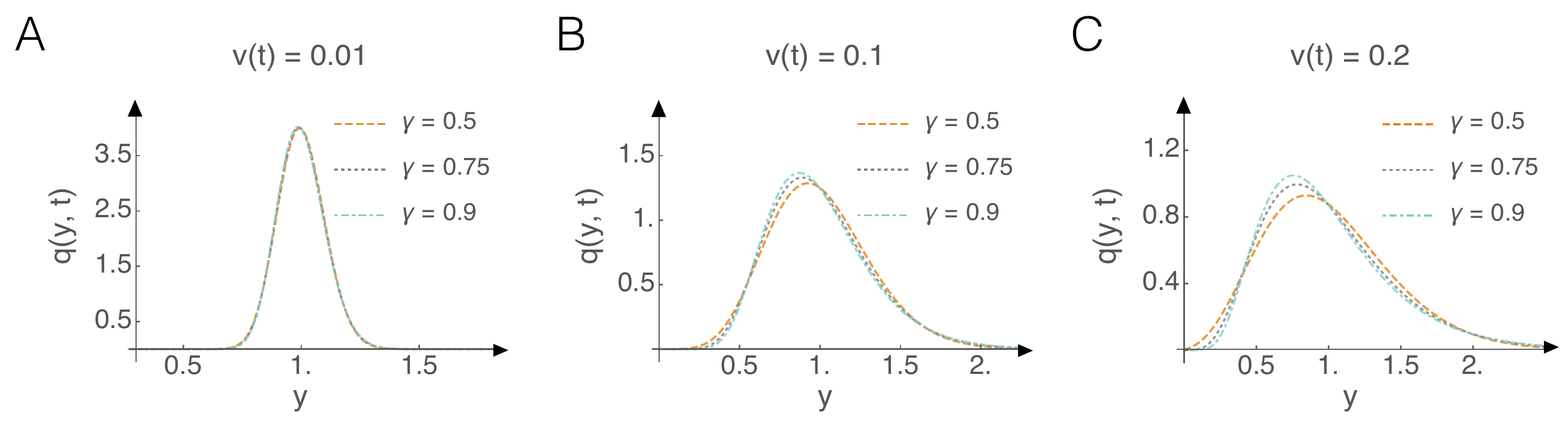}}
\caption{Plots of the mean-rescaled distribution, $q(y,t)$; it depends only on the dimensionless exponent $\gamma$, and the time-dependent dimensionless factor, $v(t) = \Sigma_\gamma^2\tau(t)$. For small $v(t)$, as shown in (A) the, dynamics are almost deterministic, and the distributions for all $\gamma$'s are well-approximated with a narrow Gaussian centered around the mean value, which is unity at all times. When the variance of the mean-rescaled distribution is kept constant (A, B and C), $\gamma$ determines its skewness.}
\label{fig:qgamma}
\end{center}
\end{figure*}

To obtain the probability density, $q(y,t)$, of the mean-rescaled variable, $y$, we start by finding an SDE for the variable $y$ using  It\^o's lemma:
\begin{eqnarray}\label{eq:ySDE}
	\der{y}{t} = \sqrt{D}\, x_0^{\gamma - 1} e^{-\kappa (1-\gamma) t} y^\gamma \xi(t),
\end{eqnarray}
with the initial condition $y_0=1$. 
Following the same approach as before,
we absorb the time dependent factor into a deterministic time change. 
We define the modified dimensionless time, $\tau \equiv \tau(t)$, such that 
\mbox{$d\tau = 2\kappa (1-\gamma)
e^{-2\kappa (1-\gamma)t}dt$}. Thus,
\begin{equation}\label{eq:taugamma}
\begin{split}
	\tau(t) &= 2\kappa(1-\gamma)\int_0^t dt\rq{} e^{-2\kappa (1-\gamma) t\rq{}} \\
		&=1 - e^{-2\kappa (1-\gamma) t}.
\end{split}
\end{equation}
$\tau(t)$ approaches $1$ as $t\to \infty$, provided that $\gamma < 1$. The time scale for the mean rescaled distribution to approach its stationary solution is set by the exponent $1/(2\kappa (1-\gamma))$. This timescale itself approaches infinity as $\gamma \to 1$, hence a stationary mean-rescaled distribution is never attained in the GBM case.

With this change of time variable \eq{ySDE} reads:
\begin{eqnarray}\label{eq:YSDE}
	\der{y}{\tau} = \Sigma_\gamma\, y^\gamma \xi(\tau).
\end{eqnarray}
\Eq{YSDE} can be reduced to a square root process with constant drift by the following change
of variable:
\begin{eqnarray}
	z \equiv y^{2(1-\gamma)}.
\end{eqnarray}
Using It\^o's lemma, one has the process that $z$ follows:
\begin{eqnarray}\label{eq:zCEV}
	\der{z}{\tau} = \Sigma_\gamma^2 \,(1-\gamma)(1-2\gamma) 
			+ 2\,\Sigma_\gamma\,(1-\gamma) \sqrt{z} \xi(\tau), 
\end{eqnarray}
with the initial condition $z_0=1$. Stochastic differential
equations with square-root multiplicative noise have been studied in detail by Feller in \cite{feller1951two};
this paper considered the general case of SDEs of the form
\begin{eqnarray}\label{eq:zFeller}
	\der{z}{t} =  b z + c + \sqrt{2a z} \xi(t).
\end{eqnarray}
This is a generalization of the square-root process with an additional constant drift term.
The process in \eq{zCEV} is obtained by the following substitutions:
\begin{equation}\label{eq:abcdef}
	a = 2\,\Sigma_\gamma^2 \,(1-\gamma)^2,\quad
	b = 0,\quad
	c = \Sigma_\gamma^2 \,(1-\gamma)(1-2\gamma).
\end{equation}
Feller \cite{feller1951two} showed that the stochastic differential equation
\eq{zFeller} has different qualitative behaviors depending on the
relative strength and sign of the coefficients. The following three cases are important to distinguish between: (i) $0<c<a$, {(ii)} $c<0$, and  (iii) $c>a>0$.

For the problem we are interested in, only two of the cases in the Feller classification
of solutions are relevant:

(i) $0 < c < a$: this corresponds to $\gamma \in (0,\frac12)$. 
The fundamental solution of the forward Kolmogorov equation corresponding
to \eq{zCEV} is not unique. There are two independent fundamental 
solutions, and the problem is well-posed only if we add an additional boundary 
condition at $z=0$, for example absorbing or reflecting boundary. {With an absorbing boundary condition, the transition density is given by \eq{zdensity} below.
}

(ii) $c < 0$: this corresponds to $\gamma \in [ \frac12, 1)$. 
There is only one fundamental solution, given by
\begin{equation}\label{eq:zdensity}
\begin{split}
	p(z,t|z_0,0) =& \frac{1}{at} \exp\le[ - \frac{z+z_0}{at}\ri]
		\left( \frac{z}{z_0} \right)^{\frac{c-a}{2a}}\\
		&\times I_{1- \frac{c}{a}}
\left(\frac{2}{at} \sqrt{z z_0}\right).
\end{split}
\end{equation}
This can be extracted from equation (1.6) of \cite{mazzon2012processo} by taking the $b\to 0$ 
limit. Substituting for $a$ and $c$ from \eq{abcdef}, we obtain 
\begin{equation}\label{eq:qgamma}
\begin{split}
	q(y,t) =& 
   \frac{y^{\frac12-2\gamma}}{\Sigma_\gamma^2 \, (1-\gamma)\,\tau(t)} 
\exp\left[ - \frac{1 + y^{2(1-\gamma)}}{2\Sigma_\gamma^2 \,(1-\gamma)^2\,\tau(t)} \right] \\
&  \times I_{\frac{1}{2(1-\gamma)}} \left( 
\frac{y^{1-\gamma}}{\Sigma_\gamma^2 \,(1-\gamma)^2\,\tau(t)} \right).
\end{split}
\end{equation}
The probability of absorption at origin is found to be 
\begin{eqnarray}
p_{\rm abs}(t) &=& 1 - \int_0^\infty q(y,t) dy \nn\\
&=& \frac{1}{\Gamma(\frac{1}{2(1-\gamma)})} \Gamma\left(
\frac{1}{2(1-\gamma)}, \frac{1}{2\,\Sigma_\gamma^2 \,\tau(t) (1-\gamma)^2} 
\right). \nn
\end{eqnarray}
{where $\Gamma(a,z)$ is the upper incomplete Gamma function, defined as
$\Gamma(a,z)=\int_z^\infty t^{a-1} e^{-t} dt$.}
The moments of the $y(t)$ can be computed in closed form using the
probability distribution function $q(y,t)$ given in \eq{qgamma}. 
As expected, the mean, $\mean{y(t)}$, evaluates to unity. The next two 
moments are 
\begin{align}
\label{eq:y2gamma}
&\mean{y^2(t)} =  \frac{1}{\sqrt{\pi}} [4a\tau(t)]^{\frac{1}{2(1-\gamma)}} e^{-\frac{1}{a \tau(t)}}\notag\\
		 &\;\;\;\;\times  \Gamma\left(\frac{2-\gamma}{2(1-\gamma)}\right)
		{}_1 F_1\left(\frac{2-\gamma}{1-\gamma}, \frac{3-2\gamma}{2(1-\gamma)},
		\frac{1}{a \tau(t)}\right),\\ \label{eq:y3gamma}
&	\mean{y^3(t)} = [a\tau(t)]^{\frac{1}{1-\gamma}} 
		e^{-\frac{1}{a \tau(t)}} 
		\frac{\Gamma\left(\frac{5-2\gamma}{2(1-\gamma)}\right)}
		{\Gamma\left(\frac{3-2\gamma}{2(1-\gamma)}\right)}\notag\\
		&\;\;\;\; \times {}_1 F_1\left(\frac{5-2\gamma}{2(1-\gamma)}, \frac{3-2\gamma}{2(1-\gamma)},
		\frac{1}{a \tau(t)}\right).
\end{align}
Here ${}_1F_1(a,b,z)$ is the confluent hypergeometric function, 
$\tau(t)$ is the function given in \eq{taugamma} and 
$a=2{\Sigma}_\gamma^2(1-\gamma)^2$. It is straightforward to show that these equations reduce to the results derived in the previous section for the $\gamma = \frac12$ case. {The moment relations \eq{y2gamma} and \eq{y3gamma} also hold for $\gamma \in (0, 1)$, provided that an absorbing boundary condition is imposed at $x = 0$.}

{\Fig{trajectories} compares the qualitative behavior (on a log-linear scale) of trajectories of the stochastic exponential growth processes described by \eq{gamma}, for different values of the two dimensionless variables which determine the dynamics, namely, $\g$ and $\Sigma_\g$. The variance of the process is set by $\Sigma_\g^2$, while the exponent $\g$ determines the relative stochasticity at short and long times. Small $\g$ values correspond to more stochastic/deterministic behavior at short/long times, while for $\g$ values around $1$, the log-scale trajectories stay stochastic over a longer time.}

\subsubsection*{The limit of small noise} 

In the large P\'eclet number limit, or small noise regime, i.e., when 
$\Sigma_\gamma \ll 1$, the distribution of $y(t)$ is sharply peaked 
around 1.  The probability density, $q(y,t)$, given by \eq{qgamma}, can be approximated using the approximation of the Bessel function for large argument:
\begin{eqnarray}
	q(y) \simeq \frac{1}{\sqrt{2\pi v(t)}} y^{-\frac32\gamma}
		\exp\left(-\frac{(y^{1-\gamma}-1)^2 }{2(1-\gamma)^2v(t)} \right),
\end{eqnarray}
in which 
\begin{equation}
	v(t)=\Sigma_\gamma^2\tau(t).
\end{equation}
The exact result for the density \eq{qgamma} depends on the model parameters
$(\kappa, D, x_0, t)$ through the same combination $v(t)$ which approximate the variance of the distribution:
\begin{eqnarray}\label{eq:varysmallsig}
	\var{y(t)} = v(t) + O\left(\Sigma_\gamma^4\,\tau^2(t)\right).
\end{eqnarray}
This is obtained from \eq{y2gamma} using the asymptotic expansion of the 
confluent hypergeometric function of large positive argument
\begin{equation}\label{eq:expansion}
\begin{split}
	{}_1 F_1(a,b;x) =& \frac{\Gamma(b)}{\Gamma(a)} e^x x^{a-b}\\
		&\times\left(1 + (-1+a)(a-b) \frac{1}{x} + O(x^{-2})\right).
\end{split}	
\end{equation}
For the small noise regime, the $t \to \infty$ limit of the variance is simply our dimensionless variance, $\Sigma_\gamma^2$:
\begin{equation}
	\lim_{t \to \infty} \var{y(t)} \approx \lim_{t \to \infty} v(t) = \Sigma_\gamma^2 \lim_{t \to \infty} \tau(t) = \Sigma_\gamma^2,
\end{equation}
which is finite for $\gamma < 1$.

\subsubsection*{It\^o vs. Stratonovich interpretations}
For {many} practical applications, the asymptotic variance of the mean-rescaled distribution, which can also be interpreted as the square of coefficient of variation of the original variable, is expected to be small\fj{~\cite{iyer2014scaling}}. Here, we show that in this region, for $\gamma < 1$, there is no difference between the behavior of \eq{gamma}, and a similar equation where the noise is interpreted in the Stratonovich sense, \ie,
\begin{equation}\label{eq:strat}
	\der{x}{t} = \k x+ \sqrt{D} x^\g \circ \xi(t).
\end{equation}
We start by writing the It\^o equivalent of \eq{strat}:
\begin{equation}
	\der{x}{t} = \k x\left(1+\frac{\g D}{2\k x^{2(1-\g)}}\right)+ \sqrt{D} x^\g \, \xi(t).
\end{equation}
Let us first examine the limiting cases: For $\g = 0$, the It\^o and 
Stratonovich equations are exactly the same. For $\g = 1$, the Stratonovich 
equation is the same as the It\^o equation with $\k \to \k + D/2$. 
For any $\g$ in between, the extra term added to the deterministic part of 
the equation decays as $x$ grows, and therefore, does not affect the dynamics 
of $x$ at long times. However, the short time dynamics can affect the 
long time distribution. This is where the small noise limit plays a role. 

From the behavior of the two limiting cases, we know that the extra term is 
more problematic as we get closer to $\g = 1$. But for $\g <1$, the additional 
term starts with a value strictly smaller than $\Sigma_\g^2$, and its expected 
value decays exponentially over time. Therefore, for $\Sigma_\g^2 \ll 1$, the 
dynamics of \eq{strat} and \eq{gamma} are the same. In summary, one does not need to worry about the choice of It\^o vs. Stratonovich schemes when phenomenologically modeling a stochastic exponential growth process.

{Let us further quantify this statement by defining $\varepsilon =  \mean{x_{\text{It\^o}}-x_{\text{Str}}}/\mean{x_{\text{It\^o}}}$ to be the relative error of choosing the incorrect prescription. Then we have
\begin{equation}
	\varepsilon = -\frac{\gamma}{2}\left(1-e^{-2\kappa(1-\gamma)t}\right)\Sigma_\gamma^2 + \mathcal{O}\left(\Sigma_\gamma^4\right).
\end{equation}
In particular, the relative error is a bounded function of time and is proportional to $\Sigma_\gamma^2$.}

{When the system} is not in the small noise limit, {It\^o and Stratonovich schemes are no longer equivalent. In this regime, we can still solve the Stratonovich problem by mapping it to an It\^o problem.} The general $\gamma$ model in Stratonovich interpretation can be reduced by a change of variable to the $\gamma=0$ case in It\^o formalism, discussed in Section III.A. If we change of variables to $y = x^{1-\gamma}/(1-\gamma)$, then we have 
\begin{align}
\der{y}{t} = \k(1-\g) y + \sqrt{D} \xi(t).
\end{align}
Now the noise is additive and there is no distinction between It\^o and Stratonovich interpretations, and so we can use the previously found solution to this model.


\subsubsection*{Possible generalizations of the models}
We comment briefly on the assumptions of our model, and possible
generalizations. The assumption of Markovian dynamics is a standard
assumption in models of stochastic kinetics \cite{VanKampen}, and simplifies
the analytical treatment. We also 
assume that the noise is modulated by the amplitude of the process being
modeled $x(t)$. A more sophisticated treatment could take into account the
stochasticity of the growth rate, $\kappa$, which could represent a 
stochastic environmental variable. A similar approach is taken in population
dynamics modeling, where the analog of $\kappa$ is an environmental variable, which might
represent factors such as food supply or temperature ~\cite{jcohen1976,tuljapurkar1980,jcohen2014}. 
One model of this type in Ref.~\cite{jmp2015} considered
a proportionate growth process where the growth multipliers
are modulated by GBM. Interestingly, in this model the growth rate of the 
mean behaves discontinuously as model parameters 
cross a critical curve, similar to what is observed in a phase transition.

\section{Statistical Inference of the Model Parameters}
Our goal in this section is to use the analytical results provided in \sect{power_law} to find a robust prescription for the determining the exponent $\gamma$ and the other model parameters, in our case $\kappa$ and $D$ from experimentally observed stochastic growth trajectories. We first, in \sect{infer_gamma}, show that there is a ratio of the third moment and the second moment of the mean rescaled distribution that is only sensitive to the exponent $\gamma$ and can be used to determine the $\gamma$. Then in \sect{infer_max}, we discuss the use of maximum likelihood estimation to find the other parameters of the stochastic dynamics. 

\subsection{Determination of $\gamma$} \label{sec:infer_gamma}

The results for the second and third moments of $y(t)$ in
\eq{y2gamma} and \eq{y3gamma} depend only on the two
combinations of variables
\begin{eqnarray}\label{eq:rgamma}
	(v(t), \gamma) = \left( 
		\Sigma_\gamma^2 \tau(t),\,
		\gamma \right).
\end{eqnarray}

Experimental measurements of the two moments $\var{y(t)}$ and
the central third moment $\mean{(y(t)-1)^3}$ can be thus translated into
a constraint for the two parameters in \eq{rgamma}.
This offers a possibility of determining the noise exponent 
parameter $\gamma$ from the shape of the $y(t)$ distribution. 

As shown in \eq{varysmallsig}, in the small-noise limit 
the variance of this distribution gives the parameter
\begin{eqnarray}
	\var{y(t)}) \simeq v(t).
\end{eqnarray}
If the time $t \gg 1/(2\kappa(1-\gamma))$ (much larger than the characteristic growth time), then the variance determines ${\Sigma}_\gamma^2 \tau_\infty = {\Sigma}_\gamma^2$.

\begin{figure}[b!]
\begin{center}
\includegraphics[width=0.48\textwidth]{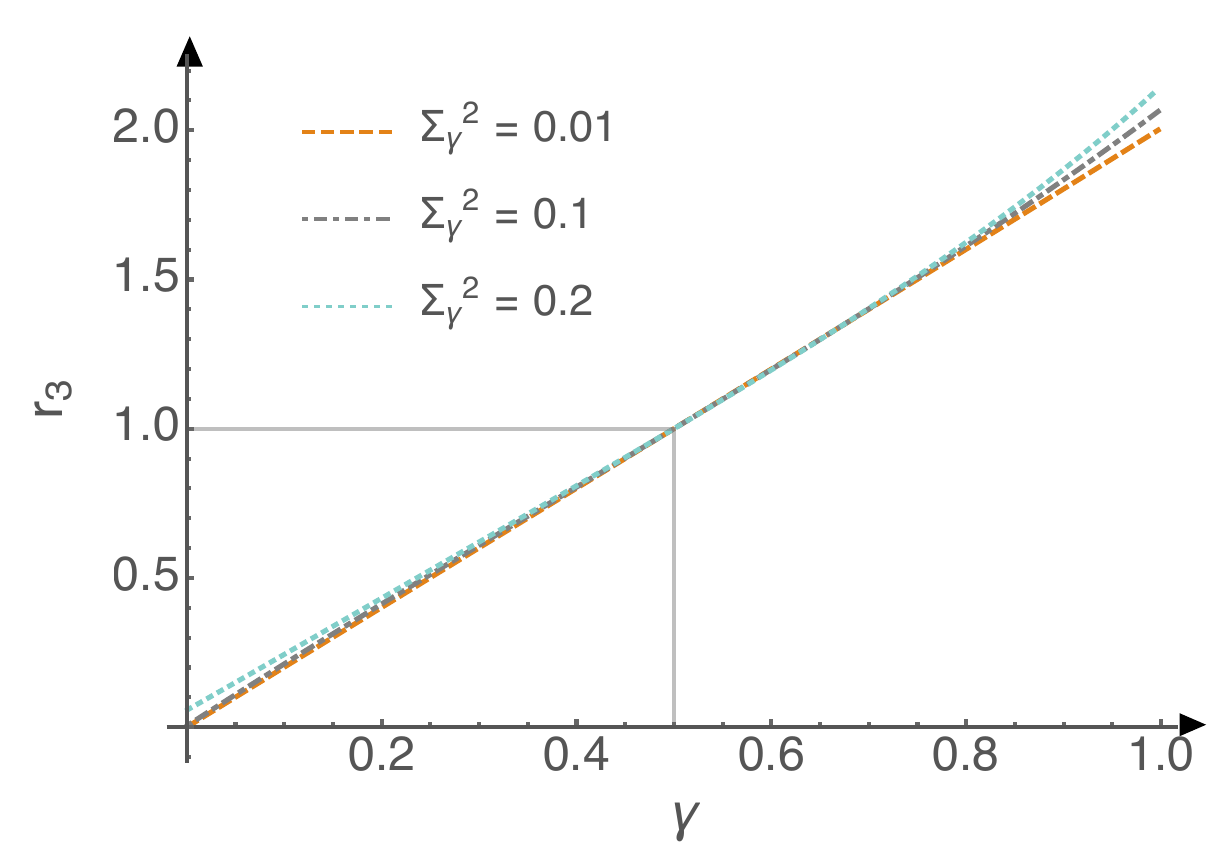}
\caption{Plots of the ratio, $r_3(\gamma,v)$, defined in \eq{r3},  as a function of $\gamma$,
	for three choices of $v = \var{y(t)} = 0.01$ (dashed, orange), 0.1 (dotted-dashed, gray)
	and 0.2 (dotted, cyan). These show that $r_3$ is sensitive to the value of $\gamma$, and changes approximately linearly with $\gamma$, but it is extremely insensitive to $\Sigma_\gamma^2$ (for small $\Sigma_\gamma^2$). As a result, $r_3$ can be used to directly deduce the value of $\gamma$ from a distribution, independent of the method of inference used to find the value of $\Sigma_\gamma^2$.}
\label{Fig:r3}
\end{center}
\end{figure}

Using \eq{qgamma} we computed the shape of the distribution $q(y)$ for 
several values of $\gamma$, for several values of the variance
$\var{y(t)}=0.01,0.1,0.2$. The results 
are shown in \fig{qgamma}, where for each case we plot $q(y)$
for $\gamma = 0.5$ (black), 0.75 (blue) and 0.9 (red).
These plots show that the shape of the distribution is sensitive to $\gamma$.

In order to determine $\gamma$ we propose the measurement of the ratio
\begin{eqnarray}\label{eq:r3}
	r_3(\gamma) = \frac23 \frac{\mean{(y(t)-1)^3}}{(\var{y(t)})^2}.
\end{eqnarray}
This is defined such that it is equal to 1 for $\gamma=1/2$, and its deviation
from 1 is a measure of the difference of $\gamma$ from $1/2$.

In Figure~\ref{Fig:r3} we show the theoretical prediction for $r_3(\gamma)$ for
three values of the variance $\var{y(t)} = 0.01, 0.1, 0.2$. 
This ratio can be computed explicitly as function of $\gamma$ and $v(t)$ as
\begin{eqnarray}\label{eq:r3explicit}
	r_3(\gamma) = \frac23 \frac{\mean{y(t)^3}-3 \mean{y(t)^2}+2}{\left(\mean{y(t)^2}-1\right)^2},
\end{eqnarray}
where the second and third moments are given explicitly in \eq{y2gamma}
and \eq{y3gamma}. The results are shown in Figure~\ref{Fig:r3}, which 
shows good discriminating value in $\gamma$, and little sensitivity to the
variance of $y(t)$. Although the theoretical results in \eq{y2gamma}
and \eq{y3gamma} are derived only for $\gamma\in [\frac12,1)$, we used them
also for $\gamma < \frac12$, in order to show that the ratio $r_3$ varies
monotonically with $\gamma$ through $\gamma = \frac12$. {As mentioned above, the theoretical results of \eq{y2gamma} and \eq{y3gamma} hold for all $\gamma \in (0, 1)$, provided that for $\gamma < 1/2$, an absorbing boundary condition is imposed at $x = 0.$}

{The insensitivity of the ratio $r_3$ to $v(t)$ at low noise limit can be understood by expanding \eq{y2gamma} and \eq{y3gamma} using \eq{expansion} in $v(t)$ and calculating $r_3$:
\begin{equation}
	r_3 = 2\gamma + \mathcal{O}\left(v(t)\right),
\end{equation}
showing why a linear relationship is observed in the small $v(t)$ limit. Note that the exact results from \eq{r3explicit} should be used to infer $\gamma$ from distributions with non-zero variance, instead of the approximate linear relationship.}

\subsection{Maximum-likelihood estimation} \label{sec:infer_max}

The method for determining $\gamma$ described in the previous section
uses only the distributional properties of the $y(t)$ variable at a fixed
time $t$. It is also possible to constrain and determine the parameters of the
stochastic process, namely, $\kappa, D,$ and $ \gamma$, by using time-series data for
the $y(t)$ process. 

The simplest estimation method for the parameters of a stochastic
diffusion is the Maximum-Likelihood Estimation (MLE) method 
\cite{bartlett1955introduction,billingsley1961statistical}. 
This has been also applied to the estimation of the parameters for a 
stochastic kinetic model~\cite{erdi2014stochastic}.

Here we provide a brief summary of the method, applied to the model at hand.
Assume that we observe the stochastic process $x(t)$ on a sequence of times
$\{ t_i\}_{i=1}^n$. 
We would like to use the time series of observations
$\{ x_{i} \}_{i=1}^n$ to determine or constrain the parameters of the
continuous-time diffusion which is assumed to have generated the observed data
\begin{eqnarray}\label{eq:target}
	\der{x}{t} = \mu(x;\theta) + \sqrt{D (x;\theta)} \xi(t)
\end{eqnarray}
The coefficients of the process depend on certain parameters $\theta$.

Define the log-likelihood function 
\begin{eqnarray}\label{eq:Ldef}
L_n(\theta) = -  \sum_{i=1}^n \log p_x(x_{i}, t_i | x_{i-1}, t_{i-1};\theta) 
\end{eqnarray}
Here $p_x(x_i,t_i|x_{i-1},t_{i-1};\theta)$ is the conditional density of $x_{i}$
given $x_{i-1}$, obtained from the diffusion \eq{target}. The log-likelihood 
function depends on the parameters $\theta$. 

The MLE method determines the parameters $\theta$ by minimizing the
log-likelihood function $L_n(\theta)$. 
This method is easy to use in cases where the conditional density $p_x$ 
appearing in \eq{Ldef} is known in closed form. This is the case for many 
popular one-dimensional diffusions, like geometric Brownian motion, the 
square root model \cite{feller1951two}, and the Ornstein-Uhlenbeck process
\cite{OUprocess}. 

Explicit results in \eq{fsqrt} and \eq{pgamma}
for the transition density in the processes \eq{sqrt}
and \eq{gamma}, respectively, given in the previous section, 
can be used to determine the parameters {$\theta =\{\kappa, D, \gamma\}$}
of these processes from time series of observations of the
stochastic variable $\{ x_i \}_{i=1}^n$. {We propose combining the
MLE method with the approach based on a measurement of $r_3(\gamma)$, 
defined in Eq.~(\ref{eq:r3}). Once $\gamma$ has been determined according to the prescription 
in Sec.~\ref{sec:infer_gamma}, the MLE will be used to infer the remaining 
parameters, $\{ \kappa, D\}$.} As an illustrative example, \app{MLE} gives a 
detailed application of this method to the estimation of the parameters of the exponential growth process with square-root noise. 

\section{Summary}
To characterize the phenomenology of stochastic exponential growth, we have 
considered a general class of {Markovian stochastic} models with power-law 
multiplicative noise described by \eq{gamma}. In this model trajectories start 
at $x=x_0$ at time $t=0$ and undergo exponential growth with rate constant 
$\kappa$ subject to power-law multiplicative noise with strength $D$ and 
exponent $\gamma$. Since the expected value of this process is a simple 
exponential growth, $x_0 e^{\kappa t}$, to understand the dynamics of the 
distribution, we only need to understand the dynamics of the distribution of 
the mean rescale variable $y= x/\mean{x}$. This paper provides an exhaustive 
study of the dynamics of {this} mean-rescaled distribution.

Starting with \eq{gamma}, we have shown that the dynamics of the 
mean-rescaled distributions  depend only on the (dimensionless) exponent 
$\gamma$, and a time-dependent dimensionless factor (see \eq{v(t)}):
\begin{equation}
v(t) = \Sigma_\gamma^2\tau(t) = \left(1-e^{-2\kappa(1-\gamma)t}\right) \frac{D x_0^{2(\gamma-1)}}{2\kappa(1-\gamma)}\,. \nn\\
\end{equation}
The latter combines the effect of all the other variables, namely 
$\kappa$, $D$, $x_0$, and, $t$. In the small noise regime, $v(t)$ is 
approximately the time-dependent variance of the distribution.  
The timescale given by $1/(2\kappa(1-\gamma))$ sets the time for the 
distribution to approach steady state. This timescale approaches infinity 
as $\gamma \to 1$, thus stationarity of the mean-rescaled distribution is 
never attained in the geometric Brownian motion (GBM) case 
(the case of $\gamma = 1$); this is a unique case for which the mean-rescaled 
distribution becomes independent of the initial condition $x_0$. On the other 
hand, for $\gamma < 1$, the mean-rescaled distributions approach stationary 
distributions with a finite variances at long times.

For $\gamma >0$, both the deterministic and the stochastic terms in 
\eq{gamma} go to zero as $x \to 0$ which seems to suggest the presence of an absorbing boundary at $x = 0$. However, due to singularity of the multiplicative noise at the origin, for $\gamma < 1/2$, the trajectories can leak through the origin, and therefore, one needs to artificially impose an additional boundary condition at the origin. The choice of boundary condition (absorbing or reflecting) affects the dynamics of the distribution in $x>0$ region. This is not the case for $\gamma \geq 1/2$, for which $x=0$ is a natural absorbing boundary.

For $\gamma < 1$, the dimensionless combination of parameters, $\Sigma_\gamma^2$, defined by $\Sigma_\gamma^2 \equiv \lim_{t\to \infty} v(t) = D x_0^{2(\gamma-1)}/(2\kappa(1-\gamma))$, sets the asymptotic variance of the mean rescaled distribution and is the analogue of the inverse P\'eclet number. When $\Sigma_\gamma \ll 1$, the dynamics are almost deterministic, and the distributions for all $\gamma$'s are well-approximated with a narrow Gaussian around the mean. However, when $\Sigma_\gamma \sim 1$ or larger, the exponent $\gamma$ determines the skewness of the distribution. We have identified a particular metric of skewness, namely, the ratio of the third central moment to the square of variance, which is sensitive to $\gamma$ and very insensitive to $\Sigma_\gamma$. Thus, this measure can be used to infer the value of $\gamma$ from experimentally measured distributions. Additionally, we have provided analytical expressions for transition probabilities $p(x,t|x_0,0)$ (the probability density of $x$ at time $t$, given $x(0) = x_0$), and a prescription that incorporates these analytical expressions to infer the remaining model parameters from  observed growth trajectories by using MLE methods.


\section{Acknowledgments}
We thank Rudro Biswas and Sasha Grosberg for insightful discussions. We acknowledge financial support from Purdue University Startup Funds and Purdue Research Foundation.

\section{Author Contributions}
SI-B conceived of and designed research;  DP, FJ and SI-B performed calculations and simulations, and wrote the paper.

\appendix
\section{Constant Noise with Absorbing Boundary: Method of Images}\label{app:constant_noise}
We start with \eq{constant}. Recalling that $x(t) = x_0 e^{\kappa t} y(t)$, the application of It\^o's lemma gives that $y(t)$ follows the diffusion:
\begin{eqnarray}\label{eq:ex1}
	\der{y}{t} = \frac{\sqrt{D}}{x_0} e^{-\kappa t} \xi(t).
\end{eqnarray}
The time dependent factor can be absorbed into the time change
\begin{eqnarray}
	\tau(t) =2 \kappa \int_0^t ds e^{-2\kappa s} = 1 - e^{-2\kappa t},
\end{eqnarray}
simplifying \eq{ex1} to
\begin{equation}\label{eq:simple_brownian}
	\der{y}{\tau} = \Sigma_0 \xi(\tau).
\end{equation}
In the absence of the boundary condition at $x=0$, the solution of the 
SDE in \eq{simple_brownian} is a time-changed Brownian motion, shifted by a constant amount. 
Its distribution is a Gaussian which is centered at $1$ and has variance 
\begin{equation}
	\var y=\Sigma_0^2 \tau(t).
\end{equation}

\begin{figure}[b!]
\begin{center}
\includegraphics[width=0.32\textwidth]{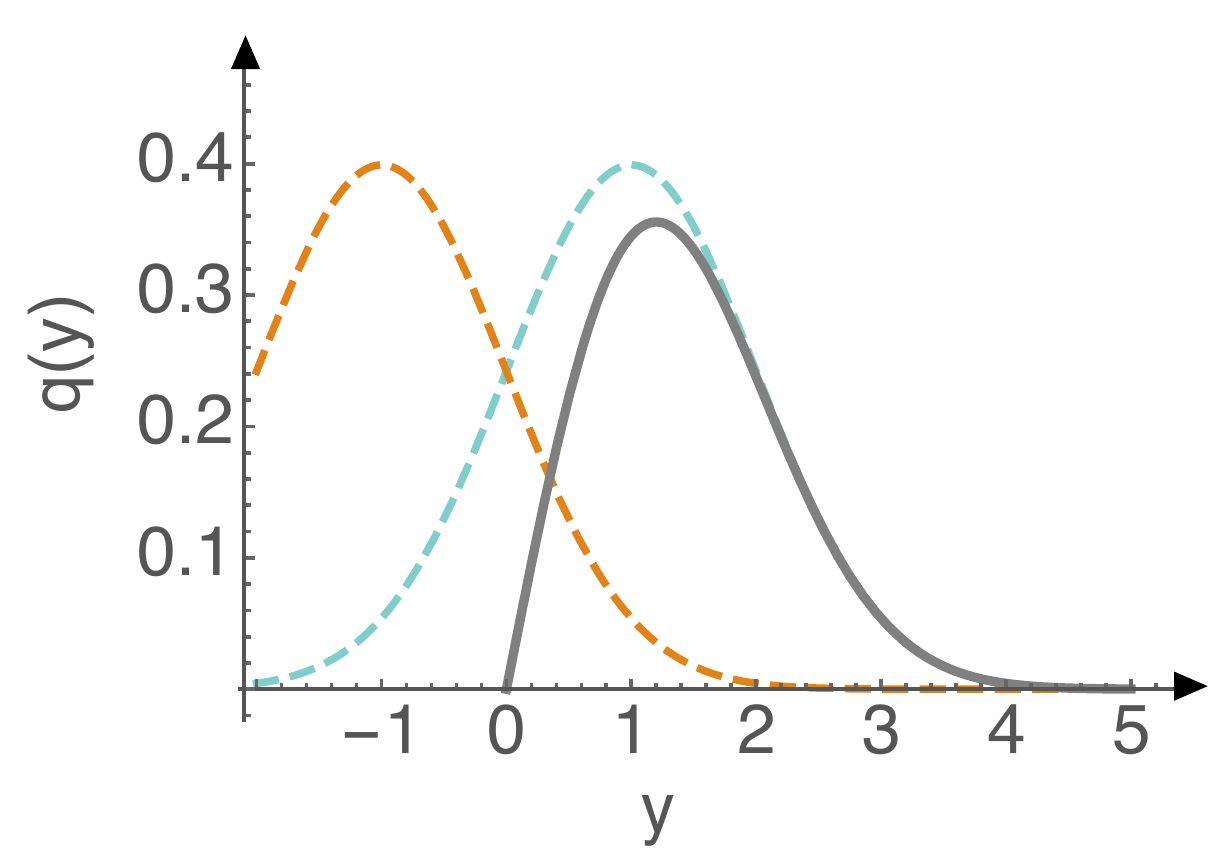}
\caption{Plot of the density $q(y)$ given in \eq{qGauss} (gray curve) for the 
stochastic exponential growth process with constant noise \eq{constant}. 
As shown in text, this can be represented as the difference of two
Gaussian distributions, centered at $y= \pm 1$ (dashed curves), see 
 \eq{qabs}.}
\label{fig:abs}
\end{center}
\end{figure}

The absorbing condition requires that the probability density of $y$ at $x = 0$, $q_{\rm abs}(0,\tau)=0$ 
for all $\tau \geq 0$. The solution of the \eq{ex1} can be found by the method
of images and can be constructed by superposition of two Gaussian
distributions, centered around $y=1$ and $y=-1$ respectively:
\begin{equation}\label{eq:qabs}
\begin{split}
	q_{\rm abs}(y,t) =&\, q(y,t;1,0) - q(y,t;-1,0) \\
			=&\, \frac{2}{\sqrt{2\pi\tau(t)} \Sigma_0} \exp
				\left( - \frac{1}{2\Sigma_0^2 \tau(t)} (y^2+1) \right)\\
			&\,\times\sinh \left(\frac{y}{\Sigma_0^2 \tau(t)}\right)\,,
\end{split}
\end{equation} 
where 
\begin{eqnarray}
	q(y,t;\pm 1, 0) =\frac{1}{\sqrt{2\pi\tau(t)} \Sigma_0} \exp
		\left( \frac{-1}{2\Sigma_0^2 \tau(t)} (y \mp 1)^2 \right).\qquad
\end{eqnarray}
It is straightforward to show that the solution, \eq{qabs}, satisfies the absorbing
boundary condition at $y=0$ for all $t\geq 0$. \Fig{abs} shows 
a plot of the density $q_{\rm abs}(x,\tau)$, and of its construction
by the method of images as the difference of two Gaussian distributions. For an alternative derivation of this result, see \cite{feller1951two}.


\section{Details of application of MLE}\label{app:MLE}

We illustrate here an example of the application of the MLE method for
determining the parameters $\kappa$ and $D$ of the process 
\begin{eqnarray}\label{eq:sqrt2}
	\der{x}{t} = \kappa\, x + \sqrt{D}\, x^{\g}\, \xi(t),
\end{eqnarray}
from a time series of observations $\{ x_i \}_{i=1}^n$ of the process 
on the sequence of times $\{ t_i \}_{i=1}^n$. \\

\begin{figure}[t!]
\begin{center}
\includegraphics[width=0.32\textwidth]{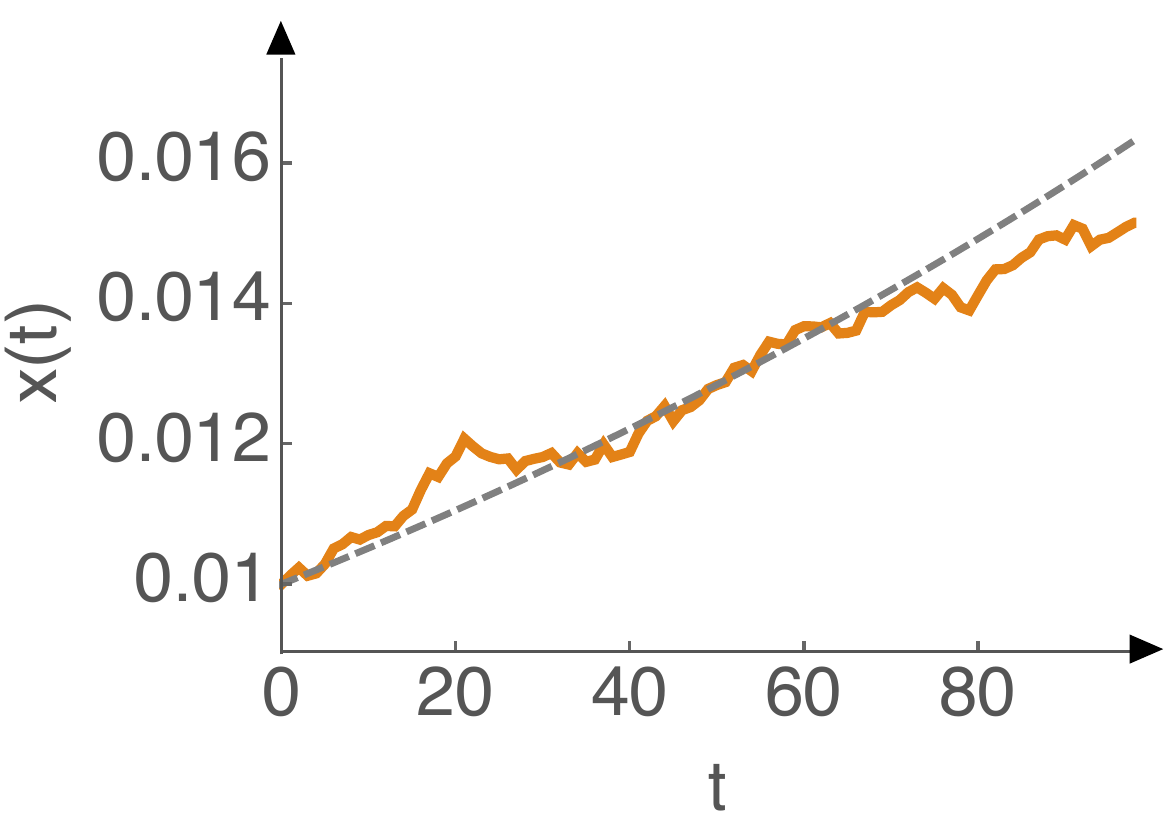}
\caption{Sample path (orange curve) of the process \eq{sqrt2} with parameters 
$\sqrt{D}=0.01$, $\kappa=0.5$, and  $x_0=0.01$. 
The dashed gray curve shows the corresponding deterministic exponential
process, $x(t) = x_0 e^{\kappa t}$.}
\label{Fig:1}
\end{center}
\end{figure}

The log-likelihood function is defined in \eq{Ldef} and requires the 
conditional transition density function $p_x(x_i,t_i|x_{i-1},t_{i-1};\theta)$. 
For the square root noise process \eq{sqrt2} this is given in \eq{fsqrt}, and
for the general $\gamma$ model it is given in (\ref{eq:pgamma}).

The MLE method determines the model parameters by minimization of the
log-likelihood function, $L_n(\theta)$, over the values of the model
parameters $\theta =\{\kappa,D\}$. A judicious choice of the initial starting point for these parameters
can speed up the minimization.

{We start by considering the simpler case of the square root process
$\gamma=\frac12$. For this case} we propose to choose the initial values 
$(\kappa_0,D_0)$ such that the first two conditional moments are matched 
exactly.  They are given by \eq{1stmom} and \eq{2ndmom}
\begin{eqnarray}\label{eq:B2}
&& \mean{x_{i+1}|x_i} = x_i e^{\kappa\tau}, \\
\label{eq:B3}
&& \mean{x_{i+1}^2|x_i} = x_i^2 e^{2\kappa\tau} +
\frac{D x_i}{\kappa} (e^{2\kappa\tau} - e^{\kappa\tau}).
\end{eqnarray}
This gives
\begin{eqnarray}\label{eq:kappa0}
\kappa_0 &=& \frac{1}{n\tau} \sum_{i=1}^{n-1} \log\frac{x_{i+1}}{x_i}, \\
\label{eq:sigma0}
D_0 &=& \frac{\kappa_0}{e^{2\kappa_0\tau}-e^{\kappa_0\tau}}
\frac{1}{n}\sum_{i=1}^{n-1} \frac{x_{i+1}^2-x_i^2 e^{2\kappa_0\tau}}{x_i}.
\end{eqnarray}

For general $\gamma \neq \frac12$, the MLE method can be reduced to the
estimation of the parameters of a square-root model by working with the
time series for $z_i = x_i^{2(1-\gamma)}$. We assume that $\gamma$ has been 
determined from a measurement of the ratio $r_3(\gamma)$, as explained
in Sec.~\ref{sec:infer_gamma}. The MLE method will be applied to determine 
the growth 
rate $\kappa$ and noise parameter $D$, for known $\gamma$.

For this case it is convenient to write the log-likelihood function in
terms of the $z_i$ variables, for which the transition density is known
from Eq.~(\ref{eq:zdensity}) and has a simpler form.

A first estimate for $\kappa$ can be obtained for general $\gamma$
again from matching the
first conditional moment (\ref{eq:B2}), which gives the estimator
$\kappa_0$ in (\ref{eq:kappa0}). The analog
of the relation (\ref{eq:B3}) is given by  
\begin{eqnarray}
\langle x_{i+1}^2 | x_i\rangle = x_i^2 e^{2\kappa \tau} 
\langle y_{\tau}^2\rangle
\end{eqnarray}
with $\langle y_{\tau}^2\rangle$ given by (\ref{eq:y2gamma}) with the 
replacement $t\to \tau$. This gives a complicated non-linear equation for $D$, 
which has to be solved to obtain $D_0$.

A simpler approach starts with the relation
\begin{eqnarray}
\langle y^{2(1-\gamma)}(t)\rangle = 1 + \Sigma_\gamma^2 (1-\gamma)(1-2\gamma)
\Delta
\end{eqnarray}
with $\Delta = 1 - e^{-2\kappa (1-\gamma)\tau}$. This
follows from the diffusion (\ref{eq:zCEV}) for $z_t = y_t^{2(1-\gamma)}$.
From this equation we obtain the conditional moment
\begin{eqnarray}
&& \langle x_{i+1}^{2(1-\gamma)} | x_i\rangle = x_i^{2(1-\gamma)} 
e^{2\kappa(1-\gamma) \tau} \\
&& \quad \times \left(1 + \Sigma_\gamma^2 \Delta (1-\gamma)(1-2\gamma)\right)
\,.\nn
\end{eqnarray}
If $\kappa \tau \ll 1$, we have to a good approximation $\Delta \simeq 
2\kappa(1-\gamma)\tau$. We obtain the following simple
estimator for $D$
\begin{eqnarray}
D_0 &=& \frac{e^{-2(1-\gamma)\kappa \tau}}{n\tau (1-\gamma)(1-2\gamma)} \\
& &\times
\sum_{i=1}^{n-1} \left( x_{i+1}^{2(1-\gamma)} - x_i^{2(1-\gamma)} e^{2(1-\gamma)\kappa\tau}
\right)\,.\nn
\end{eqnarray}
This estimator is expected to be useful when $\gamma$ is not too close to 
$\frac12$.
For this case the numerator and denominator are small, and the estimator
may be unstable numerically.

{In order to illustrate the application of this method, we generated
sample data by simulating the SDE of the square-root process $\gamma=\frac12$
using an 
Euler discretization in time. For this simulation we generated
$n=100$ values on a time grid with time step $\tau = 0.01$. The
model parameters assumed are $\kappa = 0.5, \sqrt{D}=0.01$ and
starting point $x_0=0.01$.
Figure~\ref{Fig:1} shows the sample path for this data.}

For the example data set considered here  
the initial estimators 
\eq{kappa0}, \eq{sigma0} are $\kappa_0 = 0.47749$ and
$\sqrt{D_0}=0.01918$.

Minimization of the log-likelihood function over $\kappa$ and $D$ 
with the starting point $(\kappa_0, D_0)$ gives 
$\sqrt{\hat D}= 0.01005$ and $\hat\kappa=0.47734$. 
These values are close to the actual parameter values $\sqrt{D}=0.01$
and $\kappa=0.5$. {The error of the determination decreases with
$n$, the number of the data points. 
See \cite{bartlett1955introduction,billingsley1961statistical} for 
error estimates, and the $n\to \infty$ asymptotics of the MLE errors.}


\begin{figure}[t!]
\begin{center}
\includegraphics[width=0.33\textwidth]{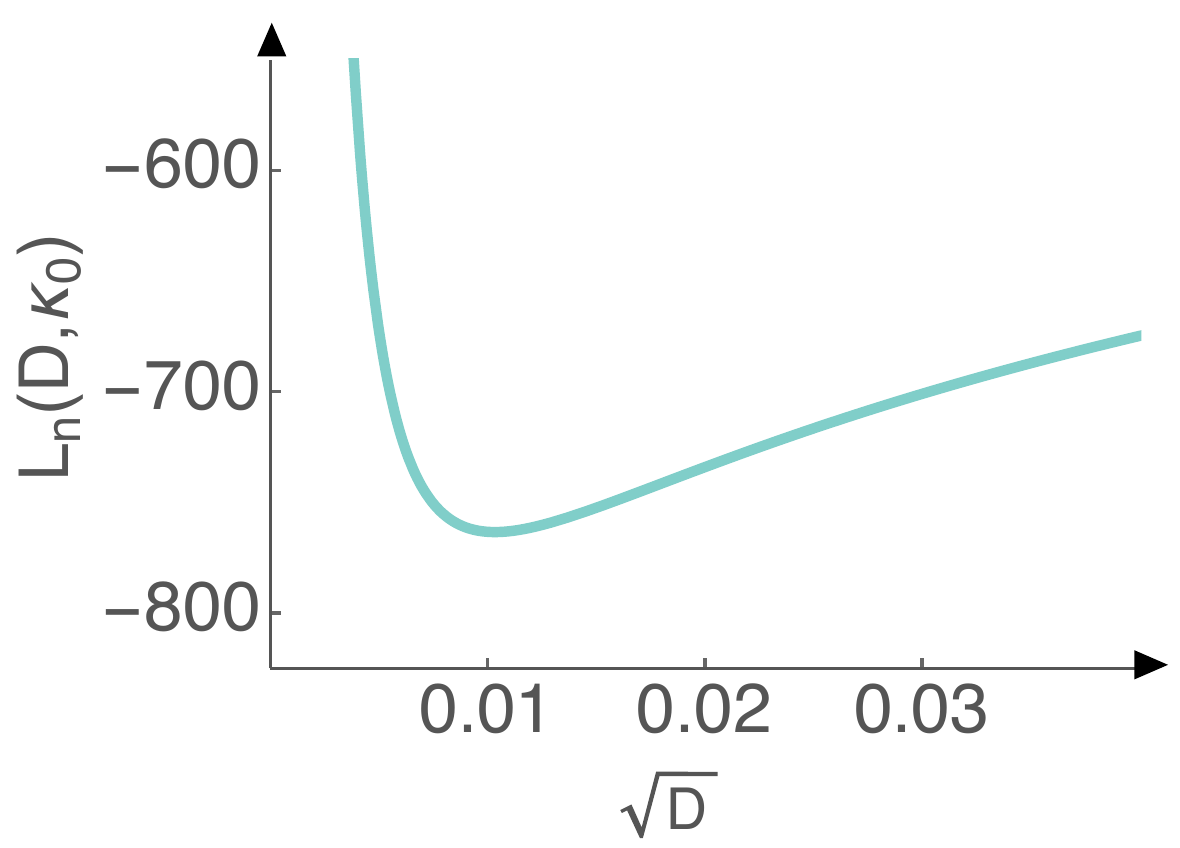}
\end{center}
\caption{
The log-likelihood function, $L_n(D, \kappa_0; \{x_i\})$, for the dataset of
Figure~\ref{Fig:1}. The $D$ parameter is determined by minimization of
this function.}
\label{Fig:2}
\end{figure}

%


\end{document}